\documentclass{emulateapj}
\usepackage{natbib}

\newcommand{\cxo}{{C}ygnus {X}-1}
\newcommand{\xray}{{X}-ray}
\newcommand{\frop}{\hbox{$\cal F$}}
\newcommand{\figref}[1]{figure~\ref{#1}}
\newcommand{\tabref}[1]{table~\ref{#1}}

\shorttitle{Shot Noise in {\cxo}}
\shortauthors{Focke, Swank, \& Wai}

\begin{document}

{.\\ \begin{flushright} \bf SLAC-PUB-11367 \end{flushright}}

\title{Time Domain Studies of {\xray} Shot Noise in {\cxo}}
\author{Warren B. Focke and Lawrence L. Wai}
\affil{Stanford Linear Accelerator Center, Stanford University}
\affil{Mail stop 98, 2575 Sand Hill Rd, Menlo Park, CA 94025}
\email{focke@slac.stanford.edu}
\email{wai@slac.stanford.edu}
\and
\author{Jean H. Swank}
\affil{X-ray Astrophysics Laboratory, NASA Goddard Space Flight Center, Code 662, Greenbelt, MD 20771}
\email{Jean.H.Swank@nasa.gov}

\begin{abstract}

We investigate the variability of {\cxo} in the context of shot moise
models, and employ a peak detection algorithm to select individual
shots.  For a long observation of the low, hard state, the
distribution of time intervals between shots is found to be consistent
with a purely random process, contrary to previous claims in the
literature.  The detected shots are fit to several model templates and
found to have a broad range of shapes.  The fitted shots have a
distribution of timescales from below 10 milliseconds to above 1
second.  The coherence of the cross spectrum of light curves of these
data in different energy bands is also studied.  The observed high
coherence implies that the transfer function between low and high
energy variability is uniform.  The uniformity of the tranfer function
implies that the observed distribution of shot widths cannot have been
acquired through Compton scattering.  Our results in combination with
other results in the literature suggest that shot luminosities are
correlated with one another.  We discuss how our experimental
methodology relates to non-linear models of variability.

\end{abstract}


\section{Introduction}

In this paper we report the results of a detailed {\xray} timing
analysis of a 97 ks {\em Rossi X-Ray Timing Explorer} (RXTE)
observation of {\cxo}.  Our time domain analysis is based upon the
phenomenological observation that the light curve is composed of
bursts of emission, called "shots."  RXTE's ability to collect long
(128-s) uninterrupted data segments greatly facilitates the shot
analysis.  Our time domain analysis is complementary to the power
spectral density (PSD) analysis of \citet{Nowak99}; we also perform an
independent cross-spectral coherence analysis with a much larger data
set.  We note that a few powerful millisecond flares have been found
in our data sample by \citet{Gierlinski03}, which are extreme examples
of the shots which we analyze here.

Shot noise has been a perennial favorite among modelers of {\cxo}'s
temporal variability.  In the simplest shot model all of the shots
have the same shape and normalization and occur at random times.  The
PSD of a realization of a simple shot model is the same as the PSD of
an individual shot, with a normalization depending on the
normalization and rate of the shots \citep{Rice54}:
$$ <P_M> = \lambda h^2 P_s, $$
where $<P_M>$ is the expectation value of the power spectrum of the
model, $P_s$ is the power spectrum of a single shot, $\lambda$ is the
average shot rate, and $h$ is the normalization of the shots.  If
there is a distribution of parameters, it may be integrated over to
obtain the resultant power spectrum. Simple exponential shot models
produce a power spectrum which is constant below a breakpoint and
$f^{-2}$ above it.  A distribution of rise or fall times which is
powerlaw between a pair of limits and zero outside them can reproduce
the observed power spectrum in the hard state.

\citet{Terrel72} used a model in which all shots had a rectangular
profile, with a timescale chosen to approximate the autocorrelation.
The difficulty of reproducing the PSD with a simple shot model in
which all shots have an identical profile has led researchers to
attempt models with distributions of shot parameters.  \citet{MiyEA88}
used a model featuring two different shot profiles to reproduce the
power spectrum and the phase lags.  \citet{LochSS91} and
\citet{BelHas90} used a model with a distribution of shot timescales
to reproduce the power spectrum.  Lochner et al. noted that almost any
shot profile could reproduce the power spectrum with a suitable
distribution of timescales, and extended this model to include a
dependence of the shot height on its timescale, and used the phase
portrait to quantify this dependence and constrain the shot profile.

\citet{NegEA94} applied a peak detection algorithm to detect
individual shots, and formed an average shot profile.  Subsequently
\citet{NegEA95} extended that investigation to include distributions
of the heights and time separations of the detected events by
tabulating the directly observed rates in the peak bins and the
separations of the peak bins of consecutive shots.  Their results were
argued by \citet{TakEA95} to support a ``sandpile'' model of self
organized criticality (SOC) in the disk, in which ``fuel'' for the
shots builds up in some location in the disk and accretion is
triggered when a critical local density is surpassed.  In this model
the shot rate is proportional to the amount of fuel present.  Shots
use up fuel, so shots are less likely to occur after another shot,
particularly a large one.

RXTE observed {\cxo} in December 1996 with its maximum effective area,
which provided the best resolution of the indivual flares (shots).
The observation took place over a period of 2 days during which the
flux, averaged over timescales longer than that typical of shots, was
within 11\% of the 2-10 keV 4600 counts $s^{-1}$. Taking this to be a
quasi-stationary "low, hard" state (based on the energy spectrum), we
examined the light curves.  We present the results in section 3, with
the implications for the SOC model.  We also investigated the energy
dependence of the variations, including the coherence; these
considerations are presented in section 4.  We discuss in section 5
how these results disagree with extended corona models.

\citet{MaccTJ00} have shown that the energy dependence of the
autocorrelation functions also disagrees with the extended corona
models. \citet{MaccTJ02} in addition carried out higher order timing
analyses on these data.  The results for the skewness implied that
there are relations between neighboring shots, that they were
correlated in arrival time and/or luminosities, so that our result on
the arrival time distribution refines this conclusion.

\citet{UttP01} pointed out that the root mean square variability (rms)
for {\cxo} and other sources was (on time scales within a state's
duration) proportional to the local mean flux level.  \citet{UttP05}
have shown that this relationship would be a natural consequence of a
particular model of non-linear processes and that it would not arise
out of a simple model of additive flux arising from independent shots.
\citet{UttP05} argue that the fundamental model should be
multiplicative processes, which then generate accumulations of flux
which are the apparent shots.  From this point of view a train of
additive independent shots does not directly reflect what is happening
in the source.  In our study the shots are a phenomenological
occurrence, whose properties we document and can compare to the
implications of a variety of models.

\section{Observations}

{\cxo} was observed by the RXTE/PCA in the hard state on 1996 December
16--17, and a couple of hours on December 15 and 18 (Observation ID
10236).  The observation spanned 182700~s of real time (29 orbits),
collecting 97311~s of data.  Two single-bit, one binned, and one event
mode were used for this observation.  The modes are summarized in
{\tabref{cx1-hard-mode-tab}}.


\begin{deluxetable}{ccccc}
\tablecaption{{\cxo} hard state guest observer data modes
\label{cx1-hard-mode-tab}}
\tablehead{\colhead{Mode} & \colhead{Type} & \colhead{$\Delta_t$ (s)} &
\colhead{$\Delta_E$ (keV)} & \colhead{\# Bins}}
\startdata
SB\_{}125us\_{}0\_{}13\_{}1s & single-bit & $2^{-13}$ & 1.5--5.0 & 1 \\
SB\_{}125us\_{}14\_{}35\_{}1s & single-bit & $2^{-13}$ & 5.0--13.1 & 1 \\
B\_{}4ms\_{}8A\_{}0\_{}35\_{}H & binned & $2^{-8}$ & 1.5--13.1 & 8 \\
E\_{}62us\_{}32M\_{}36\_{}1s & event & $2^{-14}$ & 13.1--101 & 32 \\
\enddata 
\tablecomments{{\cxo} hard state guest observer data modes.  The
``$\Delta_t$'' column contains the time resolution in s, ``$\Delta_E$'' is the
energy range covered by the mode, in keV, and ``\# Bins'' is the number of energy bins used 
by the mode.}
\end{deluxetable}


The single-bit and binned modes cover the same energy range, with
better energy resolution in the binned mode and better time resolution
in the single-bit modes.  The event mode covers the energy range from
the top of the other modes to the end of the detector's range, and is
able to provide good time and energy resolution due to the relatively
low count rates at high energies.

Background due to unrejected particles and induced radioactivity was
about 5\% of the 2-13 keV rate during the times outside the South
Atlantic Anomaly.  The modeled background, which is based on measured
particle rates during the observation, varied by 15\%. The data were
not selected on the rate of electrons entering the detectors.  It was
moderately elevated for less than 2\% of the data and there was no
excess variance or correlation of the rate of events accepted as good
{\xray} events.

\section{Shot Analysis}

\subsection{Shot Detection\label{shotdetection}}

We apply a peak detection algorithm to hard state light curves to find
shots, and examine individual events in detail.  In order to be
detected as a shot, a time bin must be the highest within a time range
to either side (the detection window, $t_d$) and must be higher than a
threshold value.  The threshold used was a multiple, $R$, of a local
average of the light curve: all points within a time range extending
equally to either side of the point in question (the filter window,
$t_f$) were averaged.  This is the convolution of the signal with a
rectangular kernel, also known as a boxcar filter.  The detection
criteria for a time bin $i$ to be accepted as a shot may thus be
expressed in closed form:
$$ \{C_j\} \leq C_i > \{C_k\};\ j \in [i-m,\ i-1],\ k \in [i+1,\ i+m] $$
and
$$ C_i > {R \over {2n+1}} \sum_{l=i-n}^{i+n} C_l, $$
where $C_x$ is the number of counts in bin $x$, and $m=t_d/\Delta_t$
and $n=t_f/\Delta_t$ are the number of time bins of width $\Delta_t$
in the detection and filter windows, respectively.  Note that the
statement that the bin is the maximum is not strictly accurate; if
multiple bins with the same rate are closer than $t_d$, the last one
and only the last one will trigger detection.  This differs from the
algorithm of \citet{NegEA94} in the definition of the local average;
they used a segmented scheme in which the local average was piecewise
constant.

\begin{figure}
\begin{center}
\includegraphics[width=0.5\textwidth]{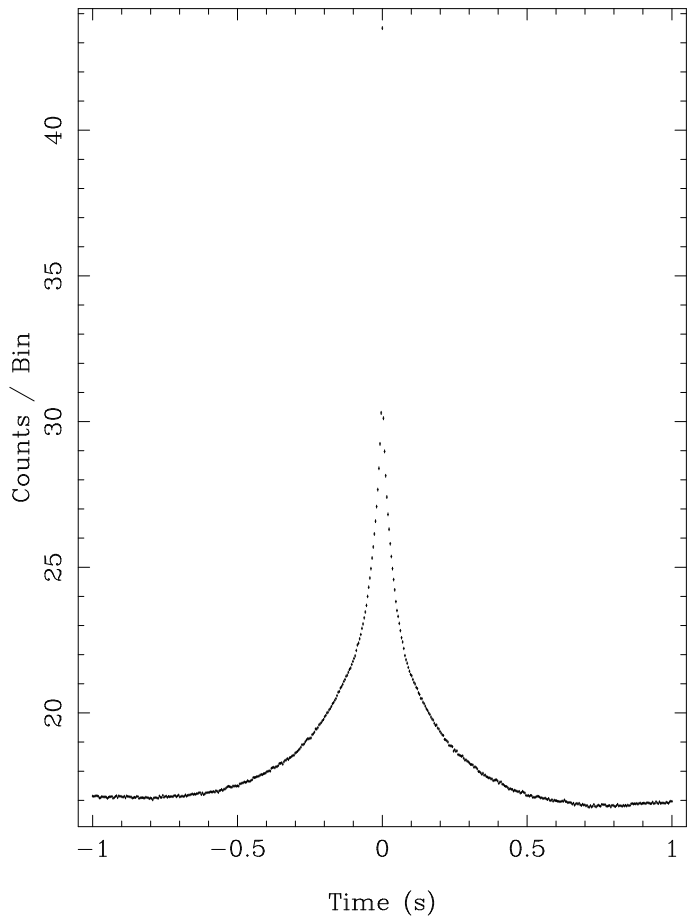}
\caption{Average shot in the hard state.  36895 2~s segments of data
surrounding the bins that triggered the detection algorithm were
averaged.  Note that the peak bin is at the top of the panel.  Its
value is artificially raised by a selection effect in the detection
routine --- since the detected peak was always a local maximum, bins
with positive Poisson fluctuations are preferentially
selected.\label{hard-avg-fig}}
\end{center}
\end{figure}

The values of the constants in the detection algorithm used were
$R=$2.0, $t_d=1.0$~s, and $t_f=4.0$~s.  97152~s of data collected by
the RXTE PCA were analyzed, yielding 36985 shots (an average rate of
0.380~s$^{-1}$), about 33000 of which were successfully analyzed.  The
data used were collected in a binned mode, with time resolution of
$\Delta_t=1/256$~s and covering an energy range of 2--13~keV.  The
average count rate was 4608~count~s$^{-1}$ and the detector dead time
was less than 2\%.  Note that at the shot peaks the deadtime is as
large as 11\%, so that the true shot profiles are actually peaked more
sharply.  In view of the larger statistical errors (about 16\% at the
peaks), the effect of deadtime was not corrected since it does not
have a significant impact on the questions being investigated.  The
average shot shape for the hard state is shown in
{\figref{hard-avg-fig}}.


This method is not without limitations.  It cannot detect events
separated by less than $t_d$, only selecting the largest events.  It
selects the highest shots, rather than the most significant --- a
short high shot would be detected preferentially to a longer, but
lower shot with higher fluence (total counts or energy).  If multiple
shots pile up, the sum is detected as a single large shot.  In fact,
since the method selects the highest points, it has a preference for
selecting times when multiple shots have piled up.  Due to statistical
fluctuations in the detected rate, it may not select the actual peak
of the shot.  This is mitigated somewhat by the process of fitting
individual shots, described below.

The method has a selection effect suppressing the detection of shots
separated by short intervals.  If two shots are closer to each other
than $t_f$ (or a little more), each raises the local average at the
position of the other, and both are less likely to be detected.  This
effect can be seen as a break in the distribution of delays between
shots.  When the detection routine was run repeatedly with different
values of $t_f$, the break point in the distribution varied
systematically, and was fairly well approximated by $t_f + t_d$.

\begin{figure}
\begin{center}
\includegraphics[width=0.5\textwidth]{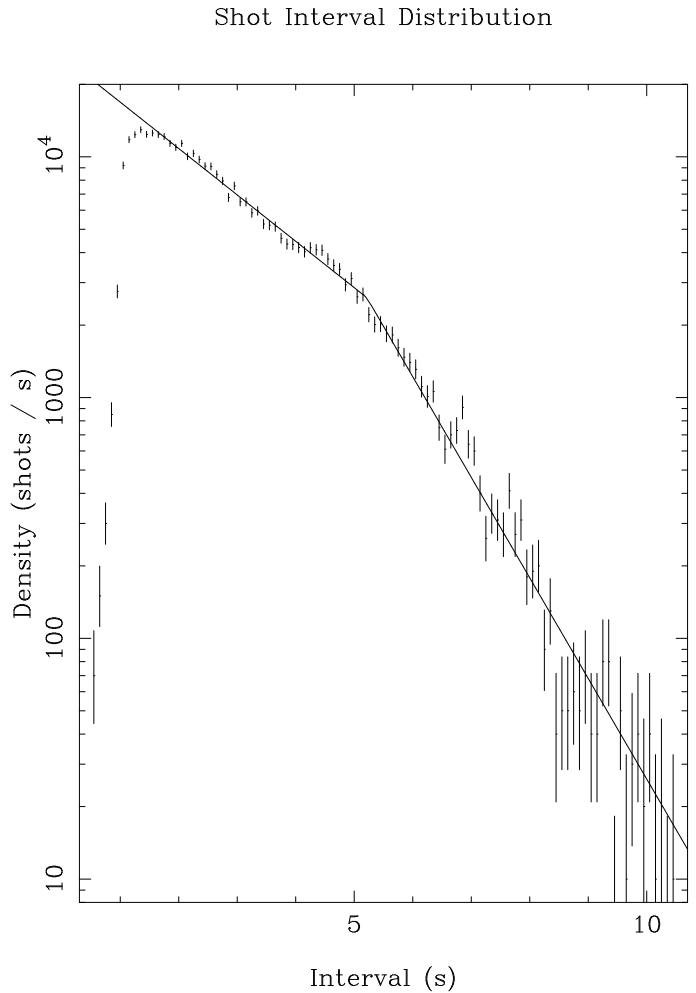} 
\caption{Distribution of time separations between shots for $t_f=4~s$,
$t_d=1~s$, and $R=2.0$.  The region above 1.5 s has been fit to a
broken exponential.  The slope above 5.2 s was $0.96s^{-1}$ and the
slope between 1.5 s and 5.2 s was $0.44s^{-1}$.  Note that the shot
peaks are obtained from the fit results described in section
\ref{shotfits}.
\label{delexp-fig}}
\end{center}
\end{figure}

If the occurrence times of the individual shots are uncorrelated (a
common assumption), an exponential distribution of the time intervals
between adjacent shots should be observed.  \citet{NegEA95} reported
that the observed distribution falls below this expected shape for
short separations.  We find that our observed distribution for
separations longer than 1.5~s (recall that the distribution is
truncated around 1~s) is consistent with an unbroken exponential.  A
break in the distribution at 5~s is expected for $t_f=4~s$ and
$t_d=1~s$.  This is shown in \figref{delexp-fig}.


\begin{figure}
\begin{center}
\includegraphics[angle=-90,width=0.5\textwidth]{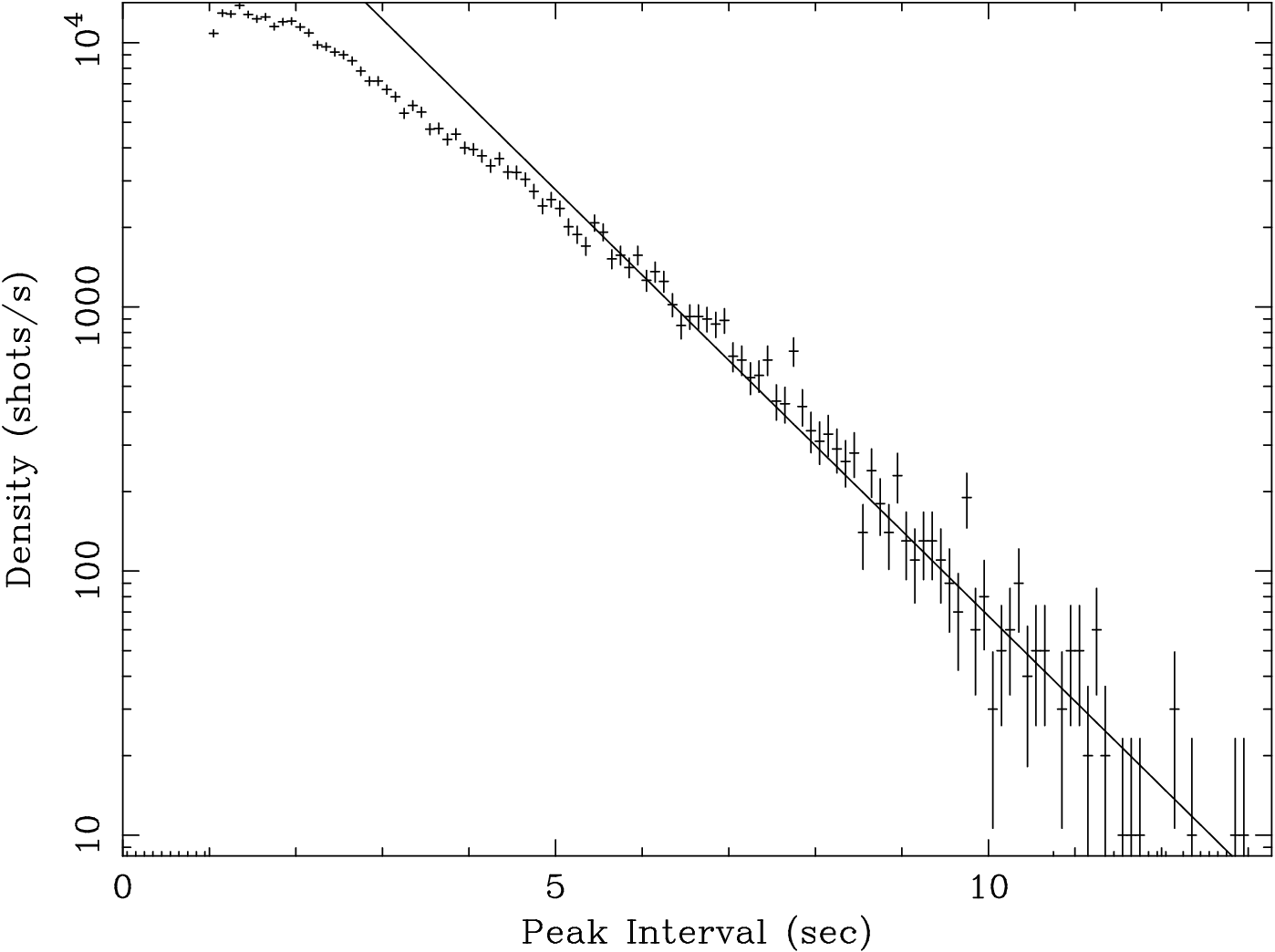} 
\caption{Distribution of time separations between shots for 8 second
intervals, $t_d=1~s$, and $R=2.0$.  The region above 5.25 s has been
fit to an exponential.  The slope above 5.25 s was $0.74 s^{-1} \pm
0.02 s^{-1}$ (90\% C.L.); also note that the slope between 2.0 s and
5.25 s (not drawn) was $0.53 s^{-1} \pm 0.02 s^{-1}$ (90\% C.L.).  The
shot peaks are defined as the peak bins from the detection algorithm
described in section \ref{shotdetection}.\label{td100_tf08-fig}}
\end{center}
\end{figure}


\begin{figure}
\begin{center}
\includegraphics[angle=-90,width=0.5\textwidth]{f4.eps} 
\caption{Distribution of time separations between shots for 32 second
intervals, $t_d=0.25~s$, and $R=2.0$.  The region above 5.25 s has
been fit to an exponential.  The slope above 5.25 s was $0.55 s^{-1}
\pm 0.02 s^{-1}$ (90\% C.L.); also note that the slope between 2.0 s
and 5.25 s (not drawn) was $0.50 s^{-1} \pm 0.02 s^{-1}$ (90\% C.L.).
The shot peaks are defined as the peak bins from the detection
algorithm described in section
\ref{shotdetection}.\label{td025_tf32-fig}}
\end{center}
\end{figure}


\begin{figure}
\begin{center}
\includegraphics[angle=-90,width=0.5\textwidth]{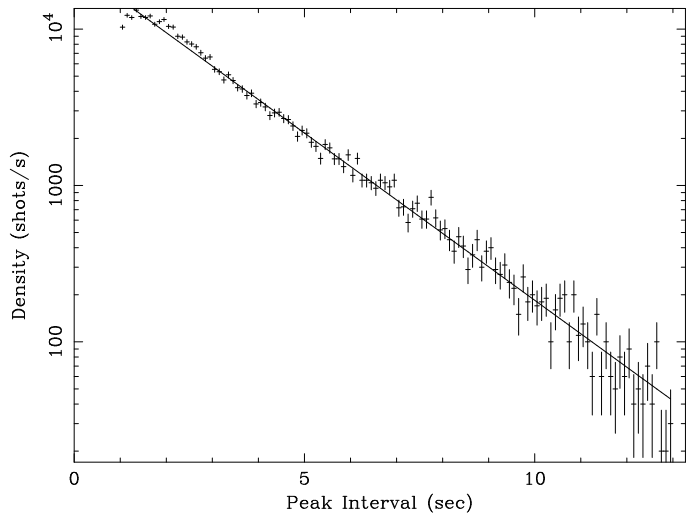} 
\caption{Distribution of time separations between shots for 32 second
intervals, $t_d=1~s$, and $R=2.0$.  The region above 5.25 s has been
fit to an exponential.  The slope above 5.25 s was $0.49 s^{-1} \pm
0.02 s^{-1}$ (90\% C.L.).  The shot peaks are defined as the peak bins
from the detection algorithm described in section
\ref{shotdetection}.\label{td100_tf32-fig}}
\end{center}
\end{figure}


To compare more directly with the \citet{NegEA95} result, we recompute
\figref{delexp-fig} with fixed 8 second piecewise constant averaging
segments to obtain the local average, and otherwise the same
parameters $t_d=1~s$ and $R=2.0$, shown in \figref{td100_tf08-fig}.
The same basic features are observed as for the boxcar filter.  Next,
we compute the case which corresponds directly to the Negoro
et. al. result from {\em Ginga} data, i.e. 32 second intervals,
$t_d=0.25~s$, and $R=2.0$, shown in \figref{td025_tf32-fig}.  For time
intervals below 1.5~seconds, we find a sharp rise in the distribution
which is consistent with the results of Negoro and can be explained by
shot pile-up.  Between 1.5 and 5 seconds we also find good agreement
with the results of Negoro et al., i.e. a systematic lowering of the
number of shots as compared to an exponential extrapolation from above
5.25 s.  Negoro et al. argue that this constitutes evidence for the
hypothesis of ``reservoirs'' in the disk or ``self-organized
criticality''.  

However, we also computed the case of 32 second intervals, $t_d$ = 1
s, and R = 2.0, shown in \figref{td100_tf32-fig}.  Raising $t_d$ to
cover the typical duration of the shots nearly eliminates the pile-up
and at the same time the depression in the range 1.5 to 5 seconds.
From comparison of \figref{td100_tf08-fig} and
\figref{td025_tf32-fig}, we see that the slope drawn in
\figref{td025_tf32-fig} is near the suppressed slope of
\figref{td100_tf08-fig}; the boundary of the supression discussed
above was moved higher when $t_f$ was raised to 32 s.  Additional
suppression is at work when $t_d$ is very short, shorter than the
duration of the shots.  Thus the suppression appears to arise from
systematic effects in identifying shots and has not provided evidence
for self-organized criticality.

\subsection{Shot Fits\label{shotfits}}

\begin{figure}
\begin{center}
\includegraphics[width=0.5\textwidth]{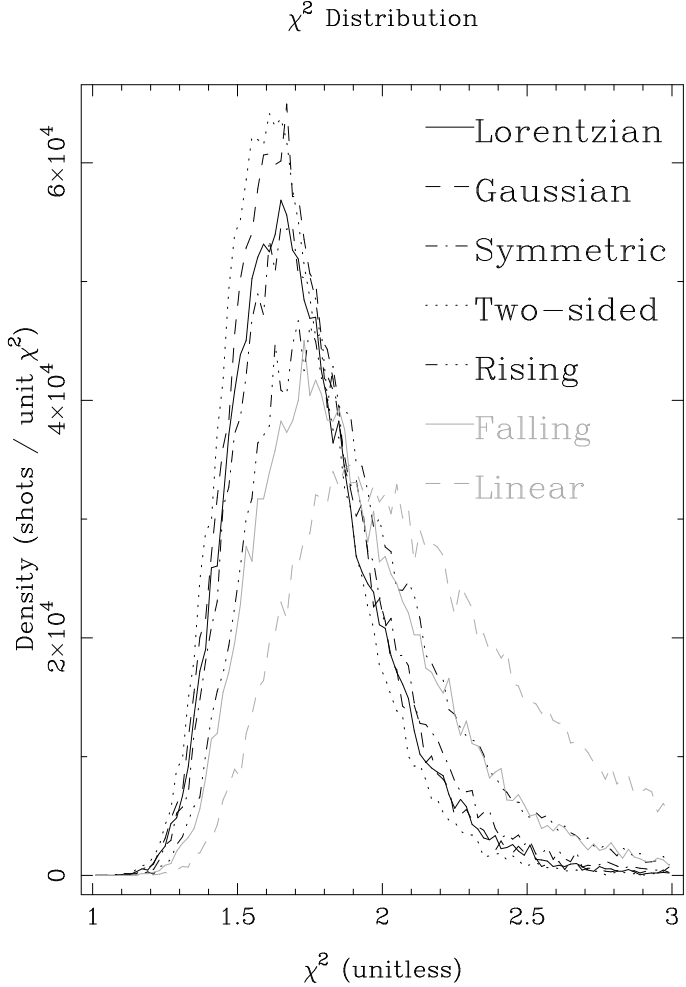}
\caption{Distributions of reduced $\chi^2$. Degrees of freedom varied
from 507 to 511 depending on the model.  The data intervals around
detected shots were fit to a linear background plus the following shot
models: A Lorentzian, a Gaussian, symmetric exponential rise and fall,
asymmetric (two-sided) exponential rise and fall, rising exponential
with a sudden fall, falling exponential with a sudden rise, and no
shot.\label{chi-fig}}
\end{center}
\end{figure}

\begin{figure}
\begin{center}
\includegraphics[width=0.5\textwidth]{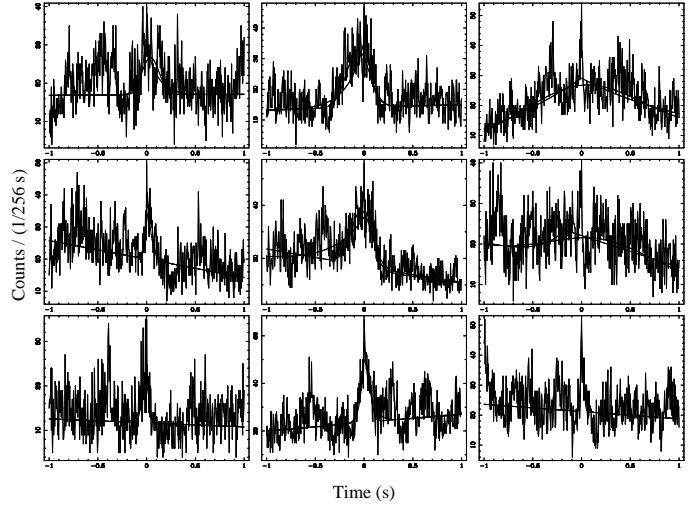} 
\caption{Several detected shots and their fits to the Gaussian and
two-sided exponential models.  Note that in two cases (the upper two
panels on the far right) a feature wider than the one that triggered
detection was fit.  In both cases, however, both models have fit the
same feature, which is taken to be a wide shot.  In the bottom right
panel, a bin near the beginning of the detection window has the same
rate as the bin which triggered detection.\label{shot-fig}}
\end{center}
\end{figure}

We employ an automated Levenberg-Marquardt fitting routine to compare
several template profiles to individual shots and report on the
distributions of the resulting parameters.  A segment of data equal to
the detection window surrounding each event was fit to the following
profiles: Lorentzian, Gaussian, symmetric exponential, two sided
exponential, rising exponential and falling exponential.  All models
included a linear background, and a simple linear model was fit to
each segment for comparison purposes.  $\chi^2$ was used as the fit
statistic rather than the C-statistic \citep{Cash79} to facilitate
comparison between models and shots.  The two-sided exponentials and
Gaussians provided the lowest $\chi^2$ values.  The distributions of
$\chi^2$ for all the models are shown in {\figref{chi-fig}}, while
several shots and their fits to the Gaussian and two-sided exponential
models are shown in {\figref{shot-fig}}.  Note that the shots were not
separated on the basis of which template fit best --- each shot
candidate was fit to all template shapes, and the distributions
associated with each template shape include all successfully analyzed
shot candidates.

 
As can be seen in {\figref{shot-fig}}, The fitting routine does not
always fit to the feature that triggered detection.  This is not
surprising: a feature which makes a small contribution to many bins
may be more significant than one which makes a larger contribution to
fewer bins.  It usually does seem to fit the same feature with the
different models.  Since we were searching for a distribution of shot
parameters, it is not disturbing that wider events are sometimes fit
--- they are the long-time tail of the shot timescale distribution.
This seeming deficiency in the fitting process may thus act to
ameliorate one of the problems with the detection process --- the
preference for detecting short, high shots over longer, lower ones
with higher significance.

\begin{figure}
\begin{center}
\includegraphics[width=0.5\textwidth]{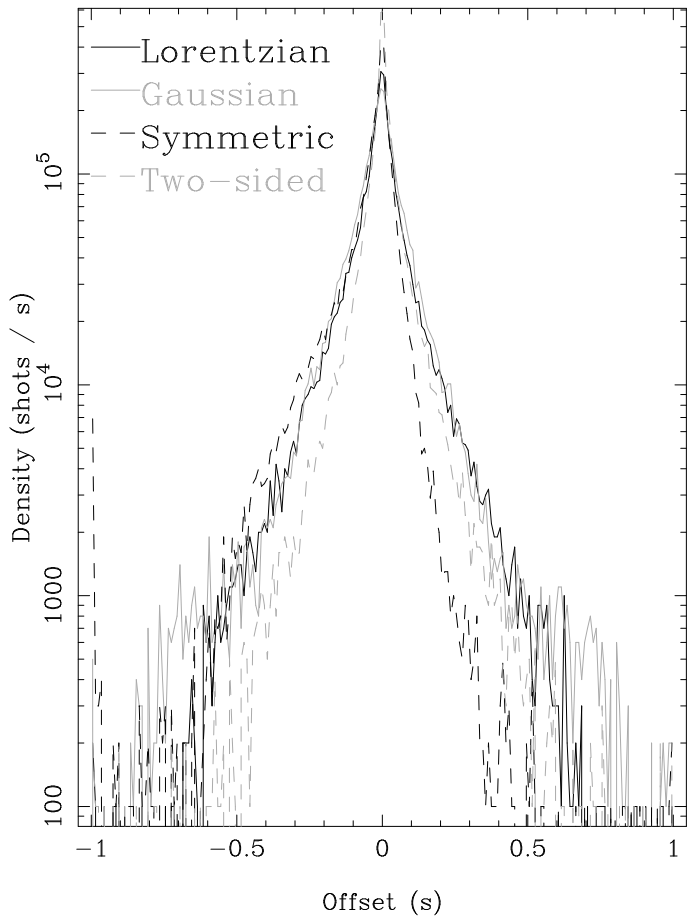} 
\caption{Distributions of offsets of the fit positions of the shots
from their initially detected position.  A marked leftward (toward
earlier times) asymmetry is visible in the wings of the distribution
for the symmetric exponential shots, and a lesser asymmetry with the
opposite sense is visible in the results for the two-sided
exponentials.  The distribution for two-sided exponentials is narrower
than the others.\label{off-fig}}
\end{center}
\end{figure}

The peak positions were allowed to vary from the initially detected
bin during the fitting, but this had little effect on the observed
distribution of time separations between shots.  This is to be
expected since the peak position could not vary outside the detection
window, and the initially detected positions cannot be closer together
than the detection window --- the observed distribution is truncated
at the lower end by $t_d$, the half-width of the detection window.
The peak positions were thus constrained to vary by less than the
minimum separation between them, resulting in little effect on the
observed distribution of separations.  While the distributions of the
offset of the fit position from the initially detected position were
all sharply peaked at zero, a marked asymmetry was visible in the
wings for the symmetric exponential shots (more were shifted to
earlier times than later), and a lesser asymmetry with the opposite
sense was observed in the two-sided exponentials.  This is shown in
\figref{off-fig}.


\begin{figure}
\begin{center}
\includegraphics[width=0.5\textwidth]{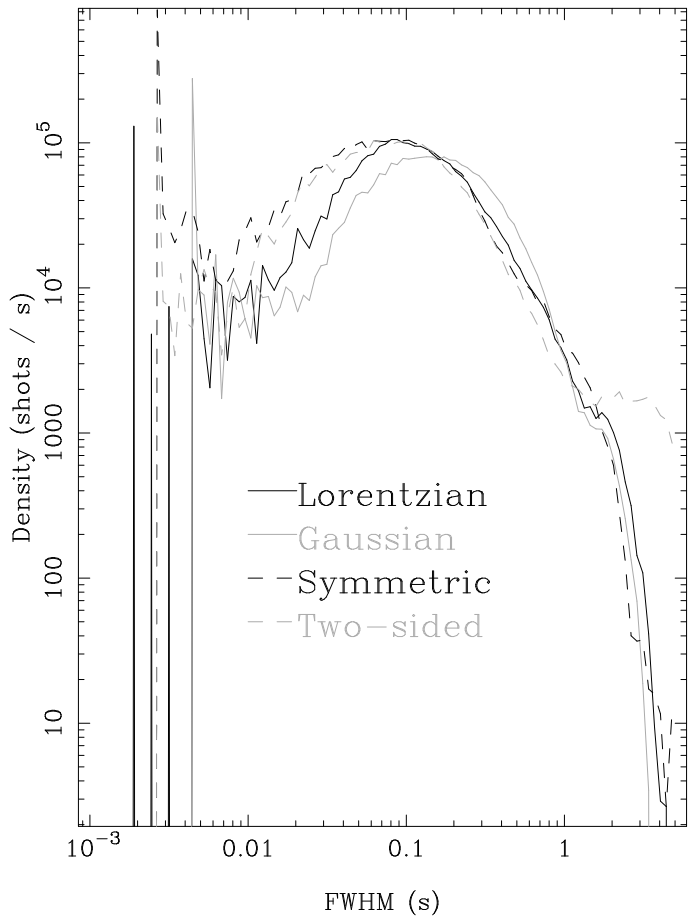} 
\caption{Distributions of shot widths.  Shots have been characterized
by their full width at half maximum.  This distribution may not show
the shortest shots which triggered the detection routine due to the
profile fitting some other feature present in the time surrounding the
shot.  Due to the bin and interval widths, times shorter than a few
tens of milliseconds or longer than a second or two may not be
meaningful.\label{fwhm-fig}}
\end{center}
\end{figure}


The distribution of shot widths is shown in \figref{fwhm-fig}.
Suggestions of very short bursts of emission \citep{RothEA77} give
interest to the number of very short shots.  The number of shots with
timescales less than 10~ms ranged from 72, when all shots are assumed
to be Lorentzian, to 327 when assumed to be symmetric exponential.
The shots with the shortest timescales may not have been characterized
well by this process --- if a shot only occupied a few bins, a greater
reduction in $\chi^2$ could be obtained by fitting some feature of the
background, which might itself be a shot with longer timescale and
lower amplitude.

\begin{figure}
\begin{center}
\includegraphics[width=0.5\textwidth]{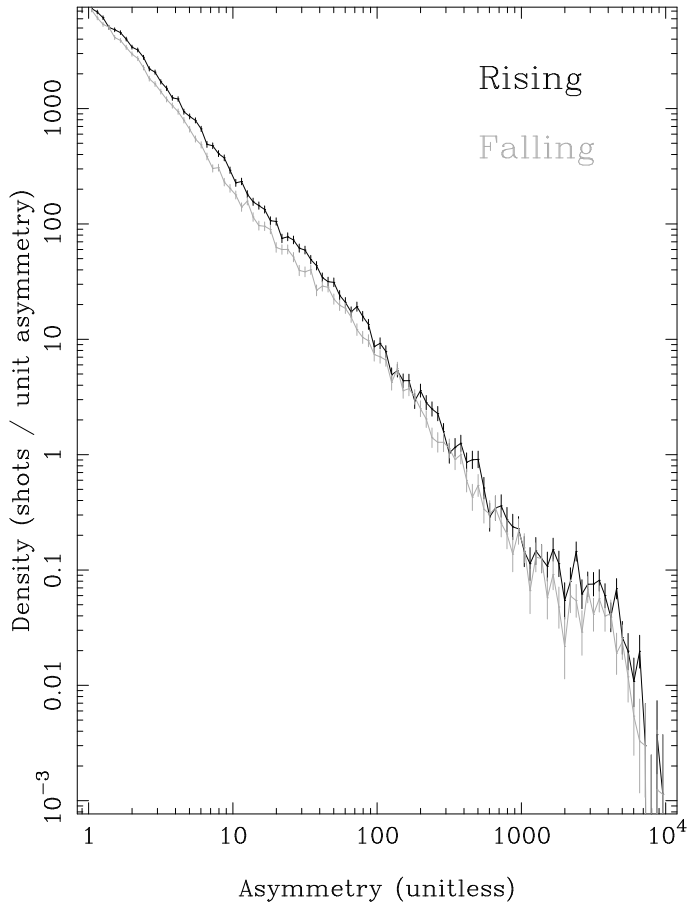} 
\caption{Distributions of asymmetry ratios for the two-sided
exponential model.  The larger of the rise or fall time was divided by
the other, and the distributions of these ratios plotted.  While the
slopes are similar, the curve for shots with a slow rise and quicker
fall (black) is higher than that for shots with a quick rise and slow
fall (gray) everywhere that the difference is significant, i.e. the
gray curve is never above the black one by more than the standard
deviation.\label{asym-fig}}
\end{center}
\end{figure}

For the two-sided exponentials, the distributions of ratios of fall to
rise or rise to fall times (whichever was larger) closely follow
similar powerlaws, slope near -1.5.  The distribution of shots with a
longer rise time is, however, higher than that for longer fall times
at all points where the difference is significant, and this is
reflected in the count of shots in each category: 18185 slow-rising
shots vs. 14611 slow-falling ones.  This can be seen in
\figref{asym-fig}.


\begin{figure}
\begin{center}
\includegraphics[width=0.5\textwidth]{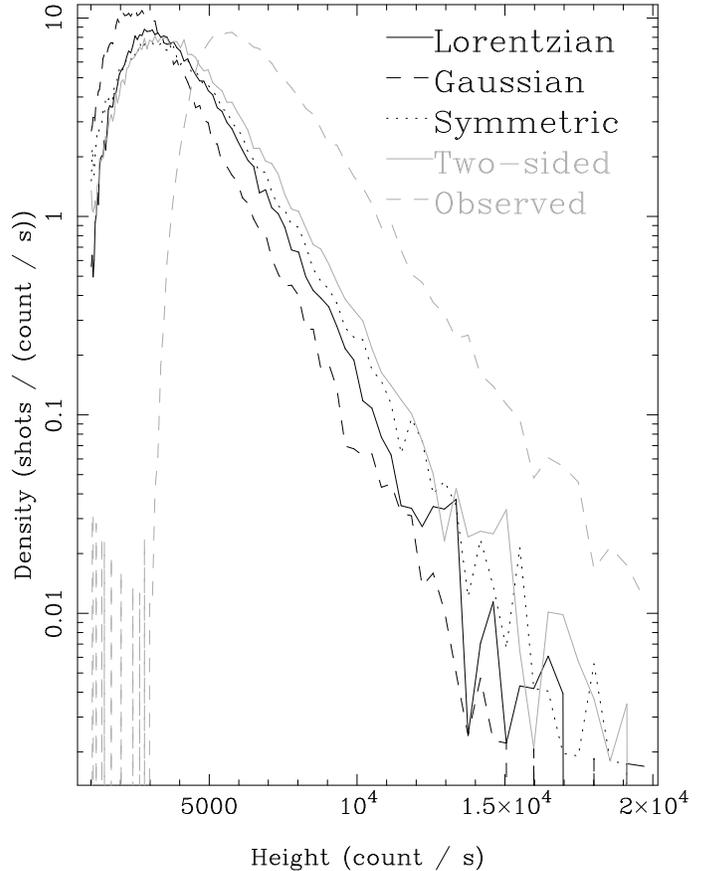} 
\caption{Shot height distributions.  Heights for 4 fit profiles are
shown, along with the heights above the local average of the bins
which triggered the detection.  This distribution is truncated at
small sizes by the detection algorithm.\label{hit-fig}}
\end{center}
\end{figure}

The distributions of fit heights for the different profiles are
similar.  This distribution is truncated at the lower end by the
requirement to be above some multiple of the local average.  These are
shown in {\figref{hit-fig}}.


The form of the distribution of directly observed rates for the peak
bin (after subtraction of the local average) is similar, but it is
displaced upwards in peak intensity when the curves are compared at
constant density levels.  This is partly due to a selection effect in
the detection criteria favoring bins with large positive statistical
fluctuations.  The separation of the curves, however, is larger than
would be expected from Poisson fluctuations alone, suggesting that the
profiles used here were not sharp enough to model the peaks of the
shots.  This interpretation is supported by the differences between
the results for the different profiles: The Gaussian profile, with the
broadest peak, is displaced furthest toward small heights, while the
exponential forms, with the narrowest peaks, are furthest toward high
peaks.  This seems inconsistent, however, with the fact that the
Gaussian profiles had lower typical $\chi^2_{\nu}$ values than any
other profile but the two-sided exponentials.  It may be that the
shots have a broad base and a narrow peak, and that the Gaussians
tended to fit the base while the exponentials tended to fit the peak.
This is also consistent with the fact that the distribution of
timescales for the Gaussian profile has more long times than those for
the exponential forms.

As we have noted, \citet{Gierlinski03} have found 4 powerful
millisecond flares in our data sample.  Their shot finding algorithm
\citep{Gierlinski03} is coarser than what we have used:
$\Delta_t=0.125$s, 128~s piecewise constant averaging segment, and
threshold defined as $10\sigma_{rms}$.  They have estimated that their
statistics are consistent with an exponential or log-normal
distribution, as extrapolated from lower amplitude shots.  This is
consistent with our analysis which shows no significant features in
the amplitude distribution at large shot heights (see
\figref{hit-fig}).

\begin{figure}
\begin{center}
\includegraphics[width=0.5\textwidth]{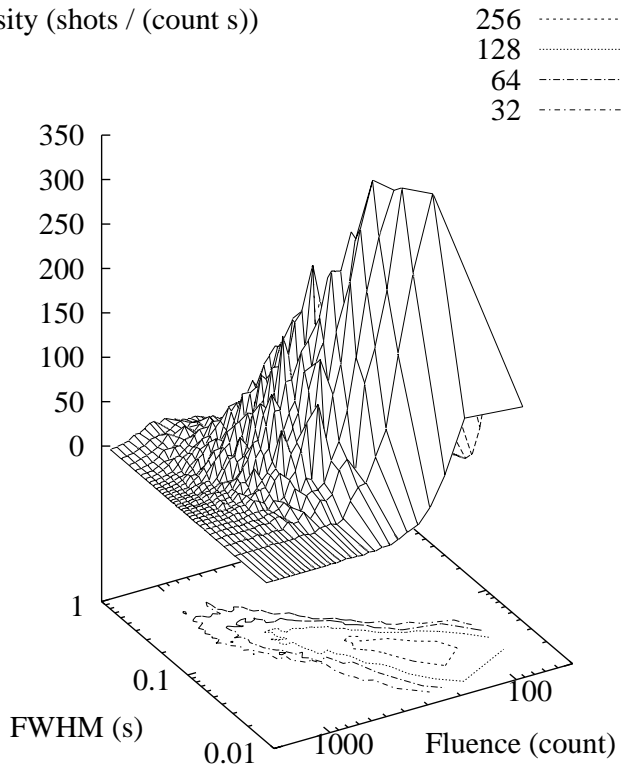} 
\caption{Joint distribution of shot timescales and fluences for
two-sided exponential shots.  The horizontal axes are shot FWHM (in
seconds) and the number of photons in the integrated shot profile.
The vertical axis is the density of shots.\label{tfjd-fig}}
\end{center}
\end{figure}

We estimate the energy by the number of photons in the integrated shot
profile (the fluence), and observe significant positive correlation in
the joint distribution of shot timescale and fluence, shown in
{\figref{tfjd-fig}}.


\section{Coherence Analysis}

The "cross spectrum" of two signals, $f(t)$ and $g(t)$, is defined by:
$$ C_{fg}(\omega) = F^*(\omega) G(\omega), $$
where $C_{fg}(\omega)$ is the cross spectrum at frequency $\omega$,
$F^*(\omega)$ is the complex conjugate of the Fourier transform of
$f(t)$, and $G(\omega)$ is the Fourier transform of $g(t)$.

The cross spectrum of two measured, binned, signals can be estimated
by segmenting them into intervals and averaging the estimated cross
spectra from each interval calculated from the discrete Fourier
transforms (DFTs) of the signals:
$$ <C_{fg,m}> = {1 \over J} \sum_{j=1}^J F^*_{m,j} G_{m,j}, $$
where $F_{m,j}$ is the DFT of signal $f(t)$ in interval $j$ (out of
$J$) in frequency bin $m$.  The cross spectrum, like the power
spectrum, may also be averaged across adjacent frequencies:
$$ <C_{fg,k}> = {1 \over m_+ - m_-} \sum_{m=m_-}^{m_+} <C_{fg,m}>. $$

The "transfer function" $h(t)$ represents $g(t)$ as linearly related
to $f(t)$ through the convolution:

$$ g(t) = h(t) \star f(t)$$

This relation may not be directly causal; for example, it may be that
both $g(t)$ and $f(t)$ are independently, causally, produced from some
unobserved "innovation signal" $i(t)$, each through its own transfer
function:

$$ f(t) = h_f(t) \star i(t), g(t) = h_g(t) \star i(t) $$

If innovation signals are applied at more than one location in the
system, the observed signals will be the sum of the two innovation
signals processed by the transfer functions for their respective
locations.  If the transfer functions for the different locations are
different, their contributions to the cross spectrum will be out of
phase and the "coherence" will be reduced.  Note that in the appendix
we perform a test of coherence loss when multiple signals are present;
we find that the simultaneous presence of different transfer functions
would result in decoherence.  Coherence is a linear correlation
coefficient between the values of the complex Fourier transforms of
the two signals as a function of frequency.  It measures the
consistency of the phase relation (that is, the consistency of
$\delta$, the phase of the transfer function) between the two signals
over time.  The coherence may be calculated as:
$$ \gamma^2 = {|<C_{fg}(\omega)>|^2 \over <|F(\omega)|^2> <|G(\omega)|^2>} $$
Note that the coherence cannot be measured with a single estimate of
the cross spectrum, since a single estimate will always be consistent
with itself.

The observed value of the coherence is reduced by the presence of
measurement noise.  \citet{VaugNow97} have derived a correction for
this effect, applicable when a good estimate of the PSD of the noise
in the signals is available.  ``Good'' here means that the uncertainty
in the estimate of the PSD of the noise is significantly less than the
PSD of the signal (which is the difference of the PSDs of the data and
the noise).  Since the main source of measurement noise in light
curves used for high-energy astrophysics is counting statistics, of
which the expected PSD and distribution thereof are well understood,
this correction may usefully be applied here.  Low coherence increases
the error on phase and coherence estimates \citep{BenPier66}.

The diagnostic utility of the cross-spectral coherence in timing
analyses is often overlooked \citep{VaugNow97}.  When the
corrections for noise are applied to the coherence between light
curves in different energy bands, the resulting value is consistent
with or close to unity.  It has been argued that this implies a
physically compact location for the source for the variability in all
energy bands.

\begin{figure}
\begin{center}
\includegraphics[width=0.5\textwidth]{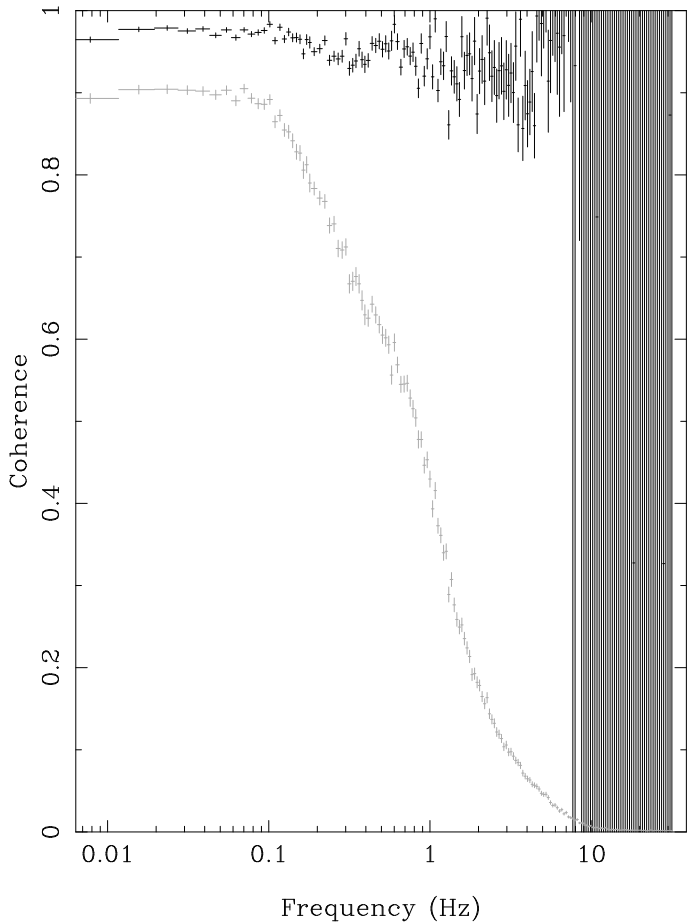} 
\caption{Hard state coherence. The coherence between the 2--3~keV and
26.5--60~keV bands in the hard state.  The black points have had the
corrections for the effects of noise applied, the grey have
not.\label{cohere-fig}}
\end{center}
\end{figure}

The cross-spectral coherence between the 2--3~keV and 26.5--60~keV
bands is shown in {\figref{cohere-fig}}.  The coherence is near unity
for frequencies below 0.1~Hz and drops to 90\% near 10~Hz; this
confirms the results obtained by \citet{Nowak99}.  At higher
frequencies, the power is dominated by noise, and coherence
corrections cannot usefully be applied.

\begin{figure}
\begin{center}
\includegraphics[width=0.5\textwidth]{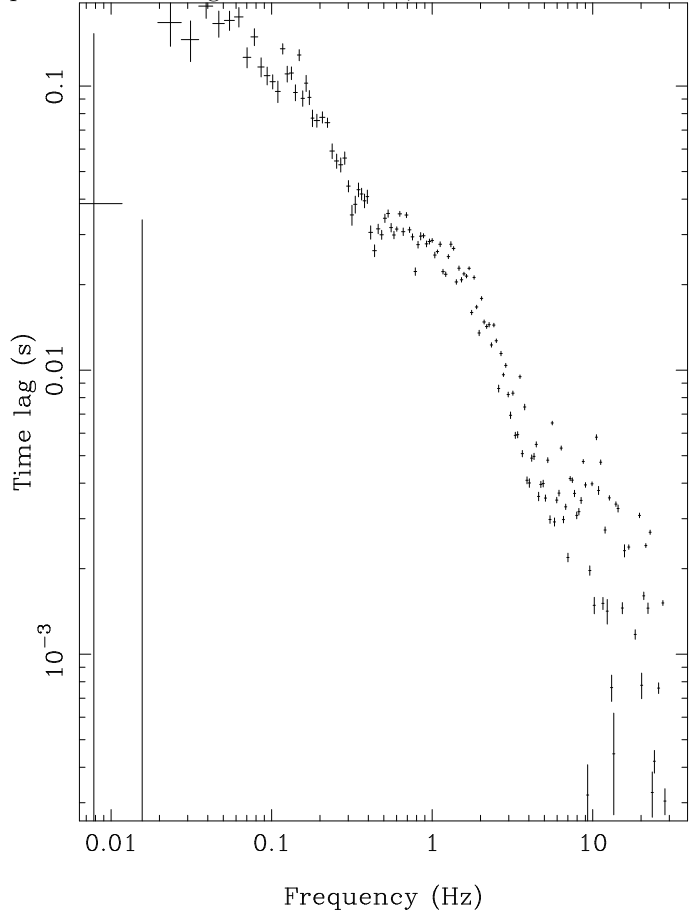} 
\caption{Hard band time lags in the hard state.  The time lags of the
26.5--60~keV band light curve relative to the 2--3~keV band calculated
with the cross-spectrum are plotted as a function of
frequency.\label{wide-time-fig}}
\end{center}
\end{figure}

Time lag here means that if $f(t)$ and $g(t)$ in the equations above
are the soft and hard band light curves, respectively, that $\delta$
is positive.  The value of the lag is a function of frequency.  The
hard band lags in the hard state are shown in
{\figref{wide-time-fig}}.  The time lag shows a power-law dependence
upon Fourier frequency, $\tau(f) \propto f^{-0.7}$, with significant
deviations from a simple power law; the longest time lag is 0.2~s.
This is comparable to the results of \citet{Nowak99}, which obtained
the same power-law dependence although with smaller longest time lag
(0.05~s).  We note that our energy band separation is larger than that
of \citet{Nowak99}, so our result confirms the observation that time
lags increase with energy band separation.


\section{Summary and Discussion}

Our main contribution here is to present the results of an analysis in
the time domain in the context of shot noise models, which is
complementary to previous analyses in the frequency domain.  Our
results on {\cxo} hard state coherence and time lags are qualitatively
consistent with the results of \citet{Nowak99} with a much larger and
independent data sample.  The notion that shot parameters such as
timescale and peak rate are distributed --- i.e., that the shots are
not all the same --- is well supported.  Models for generation of the
shots include inhomogeneity in the orbiting matter \citep{Bao95},
magnetic flares above the disk, with energy released either through
reconnection, as in solar flares, or through Comptonization of soft
radiation in the disk by the plasma trapped in the flare
\citep{NayMel98, Pout99}.  Reconnection events in the disk might also
give rise to short-term variability and hard emission.  Increased
emission due to arrival of local density enhancements at the inner
edge of the disk has also been suggested \citep{NegEA94}.
\citet{UttP05} argue that the simplest explanation of the
non-linearity that they find the data supports is that there is a
single coherent emitting region with flux from different parts of the
emitting region modulated in the same way.

The distribution of timescales constrains thermal Comptonization
models for the high-energy emission from this source.  The presence of
short shots implies that they cannot all have been processed by a
large cloud of the sort necessary to produce the observed time lags
(up to 0.2~s) between variations in different energy bands (assuming
these lags are due to Comptonization).  The presence of long shots
implies that if they have all been processed by a small corona, some
of their width must be intrinsic, and not acquired in the scattering.

There is a distribution of shot lifetimes, so if the width of the
shots is due solely to Comptonization time, they must have a
distribution of transfer functions.  But a distribution of transfer
functions should result in reduced coherence, and the observed
coherence is high --- nearly unity for all frequencies where it can be
reliably measured except for the largest energy separations available.
The presence of a broad distribution of shot timescales, along with a
high cross-spectral coherence, thus supports an intrinsic, dynamical
origin for the width of the shots.  As mentioned above,
\citet{MaccTJ02} have shown that shots should be correlated either in
arrival times and/or luminosities.  Our result of uncorrelated shot
arrival times thus implies that shot luminosities are correlated.

Within the shot picture, \citet{FengYX99a} have found that the shots
at higher energy lag behind the shock peaks in the range 2-13 keV and
that there is spectral evolution during shots.  In case there are
non-linear processes focusing towards a shot, these characteristics
will have to be shown to evolve.  Ours and other studies agree that
the characteristics of an apparent shot are intrinsic to the
occurrences there.  This is a starting point in agreement with the
non-linear picture.  It is also satisfying that whatever the origin of
the shots, their shortest time-scales are on the order of the period
that an observer at infinity should see for the innermost circular
orbit ($\approx$ 5-7 ms for {\cxo}).

\acknowledgments

This work was done at NASA's GSFC, and is from a dissertation
submitted to the Faculty of the Graduate School of the University of
Maryland, College Park by Warren B. Focke, in partial fulfillment of
the requirements for the Ph.D. degree in physics.  The revision of
this work was done at Stanford Linear Accelerator Center, Stanford
University, and supported by Department of Energy contract
DE-AC03-768SF00515.

\appendix

\section{Test of Coherence Loss when Multiple Signals are Present}

In this appendix we show the results of a Monte Carlo experiment which
tests for coherence loss when multiple signals are present.  While a
transfer function that varies in time results in reduced coherence, it
was not clear whether this would also be the case when multiple
transfer functions were present simultaneously.  That is, whether, if
the soft light curve were the sum of multiple independent signals:
$$ f(t) = \sum_{\{i\}} f_i(t),\ \frop[f(t)] = F(\omega) = \sum_{\{i\}}
F_i(\omega); $$
and the hard state were the sum of the same number of signals,
individually coherent with the corresponding soft signal, but with
different transfer functions for corresponding signal pairs:
$$ G_i(\omega) = H_i(\omega) F_i(\omega), $$
$$ g(t) = \sum_{\{i\}} g_i(t),\ \frop[g(t)] = G(\omega) = \sum_{\{i\}}
G_i(\omega) = \sum_{\{i\}} H_i(\omega) F_i(\omega), $$
$$ H_i(\omega) \neq H_j(\omega),\ i \neq j; $$
coherence would be retained.

In order to investigate this, we performed an experiment with
synthetic light curves.  Two light curves, $f_1$ and $f_2$, were
created for the ``soft'' signal, both consisting of a constant plus
one-sided exponential shots.  Time bins were the same size as in the
real data, 1/256 = 0.00390625~s.  In each curve, the constant was
1000~count~s$^{-1}$ and the shots occurred at a rate of 0.5~s$^{-1}$,
with random positions.  In one curve ($f_1$), the shots had an
$e$-folding time of 0.2~s and a height of 1000~count~s$^{-1}$, in the
other ($f_2$) they had 0.02~s and $\sqrt{10} \times 1000 =
3162$~count~s$^{-1}$ (these normalizations result in equal total power
from each set of shots).  ``Hard'' signals $g_1$ and $g_2$ were
created by delaying $f_1$ by 10 time bins and $f_2$ by 1 time bin,
respectively.  These delays are a constant fraction of the width of
the shots in the corresponding light curve.  The individual transfer
functions were thus
$$ H_i(\omega) = e^{i \omega t_i},\ t_1 = 0.0390625\ \hbox{s},\ t_2 = 0.00390625
\ \hbox{s}. $$
We used 42 intervals 8~s long to test whether the commingling of two
independent and different sets of bursts results in less coherence.
Poisson statistics were not imposed on this synthetic data.

\begin{figure}
\begin{center}
\includegraphics[width=0.5\textwidth]{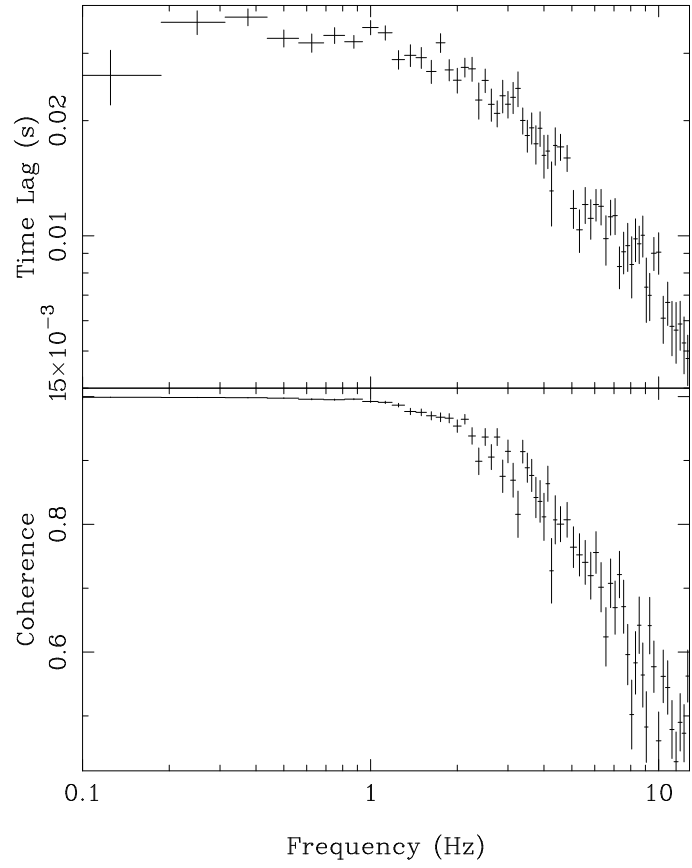}
\caption{Decoherence of simultaneous transfer functions.  The time
lags and coherence of the cross spectrum of synthetic light curves
consisting of two pairs of individually coherent functions are shown.
Note that there is no noise applied and no noise correction.
\label{fake-coh-fig}}
\end{center}
\end{figure}

These signals were added together and the cross spectrum of $f = f_1 +
f_2$ and $g = g_1 + g_2$ was calculated.  Since the phase lag of $g_1$
relative to $f_1$ reaches $\pi$ at a fraction of the Nyquist frequency
$F_c = F_{N} / 10 = 12.8$~Hz and the cross spectrum routine was not
designed with this possibility in mind, the results are only good up
to this frequency.  At low frequencies (below about 1~Hz) the cross
spectrum was coherent, with a time lag slightly below that imposed on
$g_1$.  The coherence dropped at higher frequencies, reaching 50\% at
$F_c$.  The time lag started dropping at the same frequency as the
coherence, reaching a value slightly higher than that imposed on
$g_2$.  These results are shown in \figref{fake-coh-fig}.  It should
be noted that the time delays used here are about 1/10 of those
observed in the data (to avoid further trouble with phase wrap).  This
could lessen the loss of coherence compared to what would have been
observed in the data were the observed lags due to a distribution of
transfer functions: If the phase lags are small, their difference must
be also, so the small lags used here will underestimate the loss of
coherence.  We thus conclude that the simultaneous presence of
different transfer functions would result in decoherence.


\begin{thebibliography}{}
\bibliographystyle{apj}

\bibitem[Nowak et al.(1999)]{Nowak99} Nowak, M. A., Vaughan, B. A.,
Wilms, J., Dove, J. B., \& Begelman, M.C.  1999, \apj, 510, 874

\bibitem[Gierlinski \& Zdziarski(2003)]{Gierlinski03} Gierlinski,
M. and Zdziarski, A.  2003, \mnras, 343, L84

\bibitem[Rice(1954)]{Rice54} Rice, S. O.  1954, in Selected Papers on
Noise and Stochastic Processes, ed. N. Wax (New York: Dover
Publications), 133

\bibitem[Terrel(1972)]{Terrel72} Terrel, N. J.  1972, \apj, 174, L35

\bibitem[Miyamoto et al.(1988)]{MiyEA88} Miyamoto, S., Kitamoto, S.,
Mitsuda, K., \& Dotani, T.  1988, \nat, 336, 450

\bibitem[Lochner et al.(1991)]{LochSS91} Lochner, J. C., Swank, J. H.,
\& Szymkowiak, A. E.  1991, \apj, 376, 295

\bibitem[Belloni \& Hasinger(1990)]{BelHas90} Belloni, T. \& Hasinger,
G.  1990, \aap, 227, L33

\bibitem[Negoro et al.(1994)]{NegEA94} Negoro, H., Miyamoto, S., \&
Kitamoto, S.  1994, \apj, 423, L127

\bibitem[Negoro et al.(1995)]{NegEA95} Negoro, H., Kitamoto, S.,
Takeuchi, M., \& Mineshige, S.  1995, \apj, 452, L49

\bibitem[Takeuchi, Mineshige, \& Negoro(1995)]{TakEA95} Takeuchi, M.,
Mineshige, S., \& Negoro, H.  1995, \pasj, 47, 617

\bibitem[Maccarone, Coppi, \& Poutanen(2000)]{MaccTJ00} Maccarone,
T. J., Coppi, P. S., \& Poutanen, J.  2000, \apj, 537, L107

\bibitem[Maccarone \& Coppi(2002)]{MaccTJ02} Maccarone,
T. J. \& Coppi, P. S.  2000, \mnras, 336, 817

\bibitem[Uttley \& McHardy(2001)]{UttP01} Uttley, P. \& McHardy, I. M.
2001, \mnras, 323, L26

\bibitem[Uttley, McHardy, \& Vaughan(2005)]{UttP05} Uttley, P.,
McHardy, I. M., \& Vaughan, S.  2005, \mnras, submitted
(astro-ph/0502112)

\bibitem[Cash(1979)]{Cash79} Cash, W.  1979, \apj, 228, 939

\bibitem[Rothschild et al.(1977)]{RothEA77} Rothschild, R. E., Boldt,
E. A., Holt, S. S., \& Serlemitsos, P. J.  1977, \apj, 213, 818

\bibitem[Vaughan \& Nowak(1997)]{VaugNow97} Vaughan, B. A. \& Nowak,
M. A.  1997, \apj, 474, L43

\bibitem[Bendat \& Piersol(1966)]{BenPier66} Bendat, J. S. \& Piersol,
A. G.  1966, Measurement and Analysis of Random Data, (New York: John
Wiley \& Sons)

\bibitem[Bao \& Ostgaard(1995)]{Bao95} Bao, G. \& Ostgaard, E.  1995,
\apj, 443, 54

\bibitem[Nayakshin \& Melia(1997)]{NayMel98} Nayakshin, S. \& Melia,
F. 1997, \apj, submitted (astro-ph/9710215)

\bibitem[Poutanen \& Fabian(1999)]{Pout99} Poutanen, J. \& Fabian,
A. C.  1999, \mnras, 306, L31

\bibitem[Feng, Li, \& Chen(1999)]{FengYX99a} Feng, Y. X., Li, T. P., \&
Chen L.  1999, \apj, 514, 373

\bibitem[Feng, Li, \& Chen(1999)]{FengYX99b} Feng, Y. X., Li, T. P., \& Chen L.  1999, \apj,
521, 789

\end{thebibliography}
\end{document}